\documentclass[prd,twocolumn,aps,superscriptaddress,nofootinbib,showpacs]{revtex4-1}

\usepackage[utf8]{inputenc}

\usepackage{dcolumn}
\usepackage{bm}
\usepackage{color}
\usepackage{soul,xcolor}
\usepackage{times}
\usepackage{hyperref}
\hypersetup{
  colorlinks=true,
  citecolor=blue,
  linkcolor=blue,
  urlcolor=blue}
\usepackage{epsfig}
\usepackage{dcolumn}

\usepackage{amsfonts}
\usepackage{amsmath}
\usepackage{amssymb}

\def\be{\begin{equation}}
\def\ee{\end{equation}}
\def\bea{\begin{eqnarray}}
\def\eea{\end{eqnarray}}

\def\gsim{\ \rlap{\raise 2pt\hbox{$>$}}{\lower 2pt \hbox{$\sim$}}\ }
\def\lsim{\ \rlap{\raise 2pt\hbox{$<$}}{\lower 2pt \hbox{$\sim$}}\ }
\def\dslash{\kern-4pt \not{\hbox{\kern-2pt $\partial$}}}
\def\pslash{\not{\hbox{\kern-2pt p}}}

\def\ev{{\rm eV }}

\newcommand{\dcp}{\delta_{CP}}
\newcommand{\nova}{NO$\nu$A{}}

\setstcolor{blue}

\begin{document}
\DeclareGraphicsExtensions{.eps,.ps}


\title{The Effect of a Light Sterile Neutrino at NO$\nu$A and DUNE}


\author{Shivani Gupta}
\email[Email Address: ]{shivani.gupta@adelaide.edu.au}
\homepage[\\ORCID ID: ]{http://orcid.org/0000-0003-0540-3418}
\affiliation{Center of Excellence for Particle Physics at the Terascale (CoEPP), University of Adelaide, Adelaide SA 5005, Australia}

\author{Zachary M. Matthews}
\email[Email Address: ]{zachary.matthews@adelaide.edu.au}
\homepage[\\ORCID ID: ]{http://orcid.org/0000-0001-8033-7225}
\affiliation{Center of Excellence for Particle Physics at the Terascale (CoEPP), University of Adelaide, Adelaide SA 5005, Australia}

\author{Pankaj Sharma}
\email[Email Address: ]{pankaj.sharma@adelaide.edu.au}
\homepage[\\ORCID ID: ]{http://orcid.org/0000-0003-1873-1349}
\affiliation{Center of Excellence for Particle Physics at the Terascale (CoEPP), University of Adelaide, Adelaide SA 5005, Australia}

\author{Anthony G. Williams}
\email[Email Address: ]{anthony.williams@adelaide.edu.au}
\homepage[\\ORCID ID: ]{http://orcid.org/0000-0002-1472-1592}
\affiliation{Center of Excellence for Particle Physics at the Terascale (CoEPP), University of Adelaide, Adelaide SA 5005, Australia}

\begin{abstract}
	Now that the \nova{} experiment has been running for a few years and has released some preliminary data, some constraints for the oscillation parameters can be inferred. The best fits for \nova{} include three degenerate results, the reason they are indistinct is that they produce almost degenerate probability curves. It has been postulated that these degeneracies can be resolved by running antineutrinos at \nova{} and/or combining its data with T2K. However, this degeneracy resolution power can be compromised if sterile neutrinos are present due to additional degrees of freedom that can significantly alter the oscillation probability for any of these best fits. We aim to investigate this degradation in predictive power and the effect of the DUNE experiment on it. In light of the 2018 \nova{} data we also consider the same fits but with $\theta_{23}=45^\circ$ to see if the sensitivity results are different.
\end{abstract}
\maketitle

\section{Introduction}
The existence of neutrino oscillation implies that the mass eigenstates ($\nu_1, \nu_2, \nu_3$) and flavor eigenstates ($\nu_e, \nu_\mu, \nu_\tau$) of neutrinos do not have one-to-one correspondence, instead each mass eigenstate has a different mix of each flavor eigenstate defined some mixing matrix (named the PMNS matrix after Pontecorvo, Maki, Nakagawa and Sakata). Solar, atmospheric and reactor experiments have put limits on oscillation parameters (primarily $\theta_{12},\theta_{13},\theta_{23}$ and $\Delta m^2_{21}$) 
but are unable to fully resolve the parameter space. Long baseline (LBL) experiments are required to determine some of the more elusive parameters including the mass hierarchy and the CP phase $\delta_{13}$ but unfortunately this is where several degeneracies arise.

Degeneracies are a big part of neutrino analyses due to probability expressions that contain many trigonometric terms which can, for several input parameters, output the same answer. We focus our attention the MH-$\delta_{13}$ (Mass Hierarchy-$\delta_{13}$) and Octant-$\delta_{13}$ degeneracies. When taken all together, these degeneracies imply that for certain combinations of $\theta_{23}, \Delta m^2_{31}$ and $\delta_{13}$ we will have multiple sets of parameters that give the same oscillation probability, thus an experiment may not be able to tell these situations apart. The true and test parameters we investigate therefore can be roughly divided into upper and lower ranges i.e. NH/IH (Normal Hierarchy/Inverted Hierarchy), LO/HO (Lower Octant/Higher Octant), with the midpoint between these ranges corresponding to maximal-mixing (MM).
 These ranges are defined by:
\begin{align}
\mathrm{NH}&\implies|\Delta m^2_{31}|>0,\\
\mathrm{IH}&\implies|\Delta m^2_{31}|<0,\\
\mathrm{LO}&\implies\theta_{23}<45^\circ,\\
\mathrm{HO}&\implies\theta_{23}>45^\circ,\\
\mathrm{MM}&\implies\theta_{23}=45^\circ.
\end{align}Similarly, when discussing test ranges, we also use the shorthand: WO/RO (Wrong Octant/Right Octant), WH/RH (Wrong Hierarchy/Right Hierarchy) and W$\delta_{13}$/R$\delta_{13}$ (Wrong $\delta_{13}$/Right $\delta_{13}$) to describe the test solutions surrounding the correct or incorrect regions in the parameter space.

In addition to the aforementioned parameter uncertainties, several short baseline experiments have reported results inconsistent with the three flavor oscillation paradigm, for an overview of the anomalies we refer to \cite{Gariazzo:2015rra}. A possible explanation is that oscillation is still the culprit and that this implies there is a third independent mass-squared difference which we label $\Delta m^2_{41}$. The caveat, though, is that this additional mass splitting must be much larger than the other two (roughly $1\ev$) to get such a significant effect over such short distances. Additionally this implies a fourth mass eigenstate ($\nu_4$) and hence, due to unitarity, a new flavor eigenstate ($\nu_s$) which we assume must be `sterile' to not interfere with astrophysical and particle physics constraints on the sum of active neutrino masses. Once we have this new splitting we discover that in turn we must introduce new oscillation parameters: $\theta_{14}, \theta_{24}, \theta_{34}, \delta_{14}, \delta_{34}$ and $\Delta m^2_{41}$\footnote{Note that the choice of which splitting to treat as independent and which CP phases to use is up to the physicist. For example, some papers parametrize with $\delta_{24}$ instead of $\delta_{14}$.}.

For an overview of the phenomenology and experimental constraints on a fourth neutrino we refer to Refs \cite{Dentler:2018,Aguilar-Arevalo:2018gpe,Adamson:2011ku,Abazajian:2012ys,Palazzo:2013me,Lasserre:2014ita,An:2014bik,An:2016luf,Wong:2017qob,Ade:2015xua,Gariazzo:2015rra,Gariazzo:2017fdh,Choubey:2016fpi,Aartsen:2017bap,Kopp:2013vaa,Ko:2016owz}. Similarly, for LBL analyses featuring sterile neutrinos see Refs \cite{Bhattacharya:2011ee,Hollander:2014iha,Klop:2015,Berryman:2015nua,Palazzo:2015gja,Gandhi:2015xza,Agarwalla:2016mrc,Agarwalla:2016xxa,Agarwalla:2016xlg,Dutta:2016glq,Kelly:2017kch}. For a more thorough analysis of $\theta_{23}$ and $\delta_{13}$ in the $3\nu$ case for DUNE see \cite{Srivastava:2018ser}.

For true values, we use the three best fits from The \nova{} Collaboration 2017 results \cite{Adamson:2017gxd} which are good examples of degenerate results, as well as the same results but for the MM case. Note that the significance of some of these results has dropped in the latest 2018 release \cite{NOvA:2018gge} but all are still allowed at around $2\sigma$. These solutions are outlined in TABLE \ref{ThreeSolns} with the rest of the oscillation parameters identical between each case. We aim to expand on the analyses of \cite{Goswami:2017hcw} and \cite{Ghosh2017} to analyse all three true solutions in the case where a sterile neutrino is introduced. We also produce plots with $\theta_{23}=45^\circ$ (which was previously ruled out by \nova{} but is now allowed \cite{NOvA:2018gge}) in each case to examine how the degeneracies and allowed regions change.

We will refer to these three solutions using the shorthand from TABLE \ref{ThreeSolns}. It is important to analyse these results because they are examples of solutions degenerate in probability and thus must be resolved by detector effects or combined analyses. We also analyse hypotheses with $\theta_{23}=45^\circ$ because these `maximal-mixing' solutions are allowed by MINOS, T2K and recently \nova{} at 90\% C.L. \cite{Adamson:2016xxw,Abe:2017uxa}. However, we do not fully explore the maximal-mixing parameter space because it is beyond the scope of this analysis and in general should have less issues with degeneracies.

\begin{table}[h]
	\centering
	\setlength{\extrarowheight}{0.1cm}
	\begin{tabular}{|c|c|c|c|}
		\hline
		Solution & $\delta_{13}$ & Octant & Hierarchy\\
		\hline
		A & $-90^\circ$ & LO & NH \\
		B & $135^\circ$ & HO & NH\\
		C & $-90^\circ$ & HO & IH\\
		A$'$ & $-90^\circ$ & MM & NH \\
		B$'$ & $135^\circ$ & MM & NH\\
		C$'$ & $-90^\circ$ & MM & IH\\
		\hline
	\end{tabular}
\caption{The three HO/LO and three MM true solutions considered in this analysis.\label{ThreeSolns}}
\end{table}

The main part of our analysis is introducing the sterile parameters then changing the new sterile phase $\delta_{14}$ to be several values and investigating its effect on the octant and mass hierarchy sensitivity, specifically their degeneracies. The standard three neutrino ($3\nu$) and the extended 3+1 parameters with the two representative values for $\theta_{23}$ are in TABLE~\ref{StdParam}.

\begin{table}[tbp]
	\centering
	\setlength{\extrarowheight}{0.1cm}
	\begin{tabular}{|c|c|c|}
		\hline
		$3\nu$ Parameters & True Value & Test Value Range\\
		\hline
		$\sin^2\theta_{12}$ & $0.304$ & $\mathrm{N/A}$\\
		$\sin^22\theta_{13}$ & $0.085$ & $\mathrm{N/A}$\\
		$\theta_{23}^{\mathrm{LO}}$ & $40^\circ$ & $(35^\circ,55^\circ)$\\
		$\theta_{23}^{\mathrm{HO}}$ & $50^\circ$ & $(35^\circ,55^\circ)$\\
		$\theta_{23}^{\mathrm{MM}}$ & $45^\circ$ & $(35^\circ,55^\circ)$\\
		$\delta_{13}$ & $-90^\circ,135^\circ$ & $(-180^\circ,180^\circ)$\\
		$\Delta m^2_{21}$ & $7.5\times10^{-5}\mathrm{eV}^2$ & $\mathrm{N/A}$\\
		$\Delta m^2_{31}(\mathrm{NH})$ & $2.475\times10^{-3}\mathrm{eV}^2$ & $(2.300,2.500)\times10^{-3}$\\
		$\Delta m^2_{31}(\mathrm{NH})$ & $-2.400\times10^{-3}\mathrm{eV}^2$ & $(-2.425,-2.225)\times10^{-3}$\\
		\hline
		$4\nu$ Parameters & & \\
		\hline
		$\sin^2\theta_{14}$ & $0.025$ & $\mathrm{N/A}$\\
		$\sin^2\theta_{24}$ & $0.025$ & $\mathrm{N/A}$\\
		$\theta_{34}$ & $0^\circ$ & $\mathrm{N/A}$\\
		$\delta_{14}$ &  $-90^\circ,90^\circ$ & $(-180^\circ,180^\circ)$\\
		$\delta_{34}$ & $0^\circ$ & $\mathrm{N/A}$\\
		$\Delta m^2_{41}$ & $1\mathrm{eV}^2$ & $\mathrm{N/A}$\\
		\hline
	\end{tabular}
	\caption{$3\nu$ and $4\nu$ true and test parameter values and marginalization ranges. Parameters with N/A are not marginalized over.\label{StdParam}}
\end{table}

\section{Oscillation Theory}\label{OscTheory}
Extending to $4\nu$ requires modification to the standard neutrino oscillation equations, it is important to pay attention to the parametrization chosen, because comparing mixing angles and CP phases between different choices is non-trivial. We utilize the same parametrization as in \cite{Ghosh2017}, defined as:
\begin{equation}
	U_{\mathrm{PMNS}}^{3\nu}
	=
	U(\theta_{23},0)
	U(\theta_{13},\dcp)
	U(\theta_{12},0)\,.
\end{equation}where $U(\theta_{ij},\delta_{ij})$ is a $2\times2$ mixing matrix:
\begin{equation}
	U^{2\times 2}(\theta_{ij},\delta_{ij})
	=
	\left(
	\begin{array}{c c}
		\mathrm{c}_{ij} & \mathrm{s}_{ij}e^{i\delta_{ij}}\\
		-\mathrm{s}_{ij}e^{i\delta_{ij}} & \mathrm{c}_{ij}
	\end{array}
	\right)
\end{equation} in the $i,j$ sub-block of an $n\times n$ identity array with trigonometric terms abbreviated with the notation:
\begin{align}
	\mathrm{s}_{ij}=&\sin\theta_{ij},\\
	\mathrm{c}_{ij}=&\cos\theta_{ij}.
\end{align}The four flavor parametrization is then:
\begin{equation}
	U_{\mathrm{PMNS}}^{4\nu}=
	U(\theta_{34},\delta_{34})
	U(\theta_{24},0)
	U(\theta_{14},\delta_{14})
	U_{\mathrm{PMNS}}^{3\nu}
\end{equation}With new mixing angles: $\theta_{14},\theta_{24},\theta_{34}$ and phases: $\delta_{14},\delta_{34}$. The fourth independent is chosen to be $\Delta m^2_{41}$ for consistency.

The probability expression is simplified with approximations as detailed in \cite{Klop:2015,Ghosh2017}. Note that the $\Delta m^2_{41}$ terms are averaged over to represent the limited detector resolution, removing explicit dependence, leaving:
\begin{align}\label{Pnue}
	P^{4\nu}_{\mu e}
	=
	&\quad (1-\mathrm{s}^2_{14}-\mathrm{s}^2_{24})\Big[4\mathrm{s}^2_{23}\mathrm{s}^2_{13}\sin^2\Delta_{31}\\ 
	&+8\mathrm{s}_{13}\mathrm{s}_{12}\mathrm{c}_{12}\mathrm{s}_{23}\mathrm{c}_{23}\sin\Delta_{21}\sin\Delta_{31}\cos(\Delta_{31}+\delta_{13})\Big]\notag\\
	&+4\mathrm{s}_{14}\mathrm{s}_{24}\mathrm{s}_{13}\mathrm{s}_{23}\sin\Delta_{31}\sin(\Delta_{31}+\delta_{13}-\delta_{14})\notag.
\end{align}where:
\begin{align}
\Delta_{ij}=&\frac{\Delta m^2_{ij}L}{4E}.
\end{align} The $\Delta_{31}, \delta_{13}$ and $\delta_{14}$ dependent terms can lead to the MH-CP degeneracies, due to the unconstrained\footnote{The entire range of $\delta_{13}$ is allowed at $2\sigma$ for NH, while in IH most of the range is allowed at $3\sigma$, though the approximate 1/4 plane centred on $\delta_{13}=90^\circ$ is excluded.} CP phases ($\delta_{13}$ and $\delta_{14}$) and sign of $\Delta_{31}$. Note also that the antineutrino probability can be obtained by performing the replacements: $\delta_{13}\rightarrow-\delta_{13}$ and $\delta_{14}\rightarrow-\delta_{14}$.

\section{Experiment Specification}
We run our simulation for the currently running \nova{} experiment \cite{Ayres:2004js,Messier:2013sfa} (with modified experimental set-up taken from Ref. \cite{Agarwalla:2012bv}) as well as the future experiment DUNE \cite{Acciarri:2016crz,Alion:2016uaj}. To simulate these experiments we use the GLoBES package along with auxiliary files to facilitate sterile neutrino simulation \cite{Huber:2004ka,Huber:2007ji,Huber:2007xx2,Huber:2009xx}.

\nova{} is a USA based experiment with a baseline of 812 km. It runs from Fermilab's NuMI complex in Illinois to a far detector in Ash River Minnesota. We assume that \nova{} will run for a total of three years in neutrino mode and three years in antineutrino mode ($3+\bar{3}$).

If these degeneracies can be solved at all with the current experiments T2K \cite{Abe:2017uxa} and \nova{} \cite{Adamson:2017gxd} then they may give the first hints of the values of $\delta_{13}$, $\theta_{23}$ and the sign of $\Delta m^2_{31}$ at some significant confidence level.

The addition of sterile neutrinos to the oscillation model can greatly lower sensitivity to degeneracies for \nova{} and T2K \cite{Agarwalla:2016mrc}, and DUNE is already predicted to have very good degeneracy resolution \cite{Acciarri:2015uup,Ghosh:2014rna}
for $3\nu$ so it's important to see how much it's affected by the sterile neutrino. In addition, to see the how the sensitivity scales for runtime, we simulate DUNE for $2+\bar{2}$ and $5+\bar{5}$.

It is predicted that DUNE, along with other proposed next generation long-baseline experiments such as T2HK (Tokai to Hyper-Kamiokande) \cite{Abe:2011ts} and/or T2HKK (Tokai to Hyper-Kamiokande and Korea) \cite{Abe:2016ero} will be very sensitive to sterile induced CP phases \cite{Choubey:2017cba,Choubey:2017ppj}. As such, they will contribute much further to oscillation physics once the current degeneracies and issues are resolved, especially if sterile neutrinos are present.

\section{Identifying degeneracies in the 3+1 case}
\label{IdentifyingDegens}

\subsection{Degeneracies at the probability level}
After taking the standard best fits for oscillation parameters from sources such as global fits and oscillation experiments \cite{Forero:2014bxa,Esteban:2016qun,Capozzi:2013csa} and choose sterile parameters consistent with \cite{Kopp:2013vaa,Giunti:2011gz,Giunti:2013aea,Gariazzo:2017fdh} we then set $\theta_{34}$ and $\delta_{34}$ to zero because they are not present in the vacuum equation for $P_{\mu e}$, Eq. (\ref{Pnue}), and we are under the assumption that matter interactions will not add any significant dependence to these terms.
Finally we smooth our curves with a moving box-windowed average to represent the small oscillations that will be present but cannot be seen in real data, as mentioned in section \ref{OscTheory}. 

When we plot the probability plots for our three true values into the $4\nu$ sector and vary $\delta_{14}$ from $-90^\circ$ to $+90^\circ$, our lines will become bands. This may cause additional overlap where there was none before, thus introducing or re-introducing specific degenerate solutions. This is the primary feature we are interested in as it will determine the sensitivity degradation that would be present in the $3+1$ case.

 For the plots where they are not axis variables we marginalize $|\Delta m^2_{31}|$, $\delta_{13}^{\mathrm{test}}$ and $\delta_{14}^{\mathrm{test}}$ to minimize $\chi^2$ in the fit. All of the marginalization ranges are summarized in TABLE~\ref{StdParam}.
 
 It can be seen from FIG \ref{NOvAprob3nuMaxMix} that the curve separation for antineutrinos relative to the neutrino case seen in HO/LO is lessened for MM. This Implies that it'll be less important to run antineutrinos to distinguish these three values. This is due to the octant-$\delta_{13}$ degeneracy vanishing as $\theta_{23}$ approaches $45^\circ$. The MH degeneracy for results B$'$ and C$'$ is still significant in all cases as with B and C.
 
\subsubsection{NO$\nu$A}
\begin{figure}[h]
	\begin{tabular}{cc}
		\includegraphics[width=0.25\textwidth]{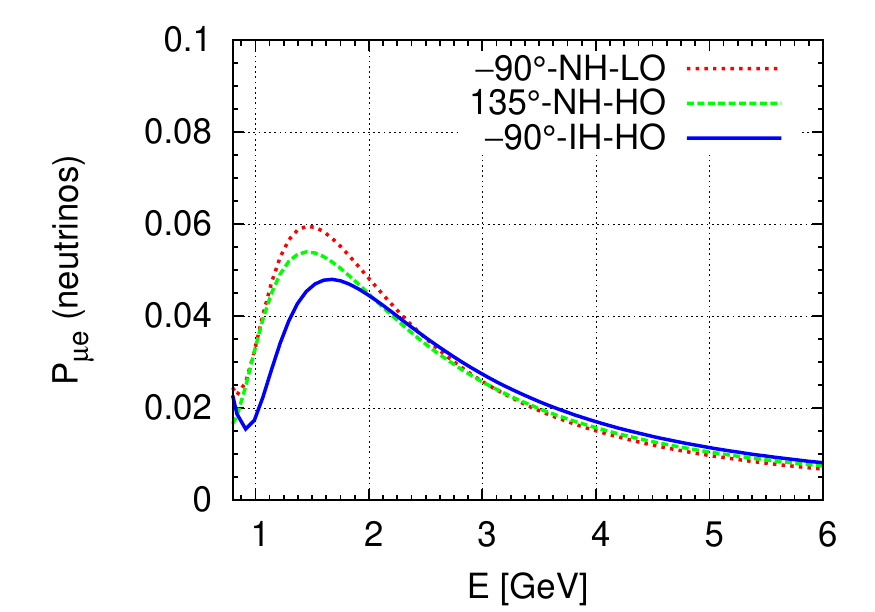}
		&
		\includegraphics[width=0.25\textwidth]{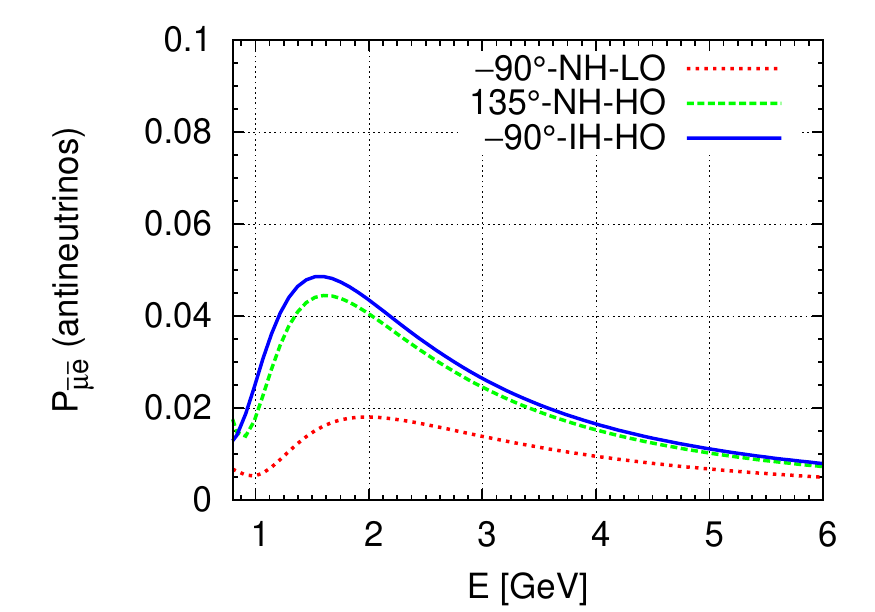}
		\\
		(a) Neutrinos.
		&
		(b) Antineutrinos.
	\end{tabular}
	\caption{Three-flavor probability plots with all three true value lines overlaid for \nova{} showing the largely degenerate curves except in the antineutrino case where the LO curve is distinct.\label{NOvAprob3nu}}
\end{figure}

\begin{figure}[h]
	\begin{tabular}{cc}
		\includegraphics[width=0.25\textwidth]{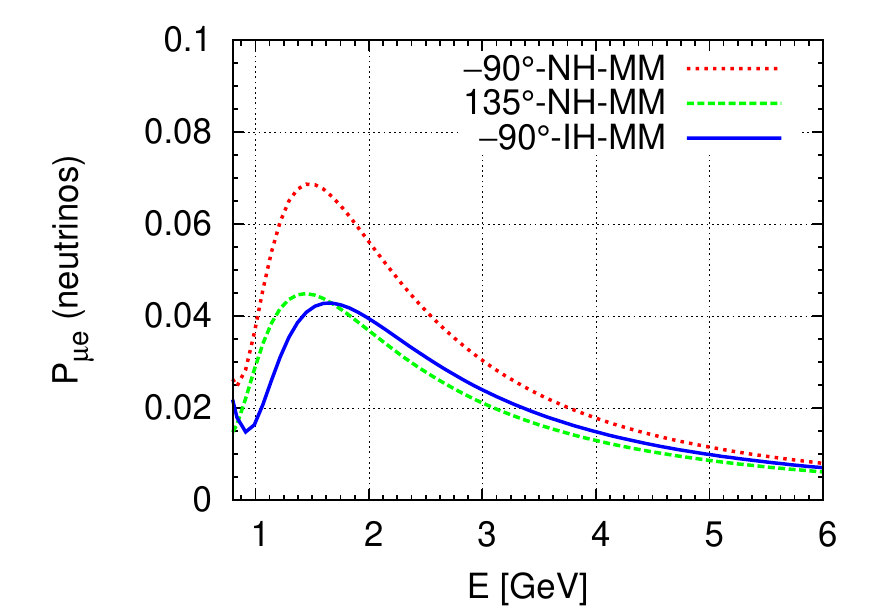}
		&
		\includegraphics[width=0.25\textwidth]{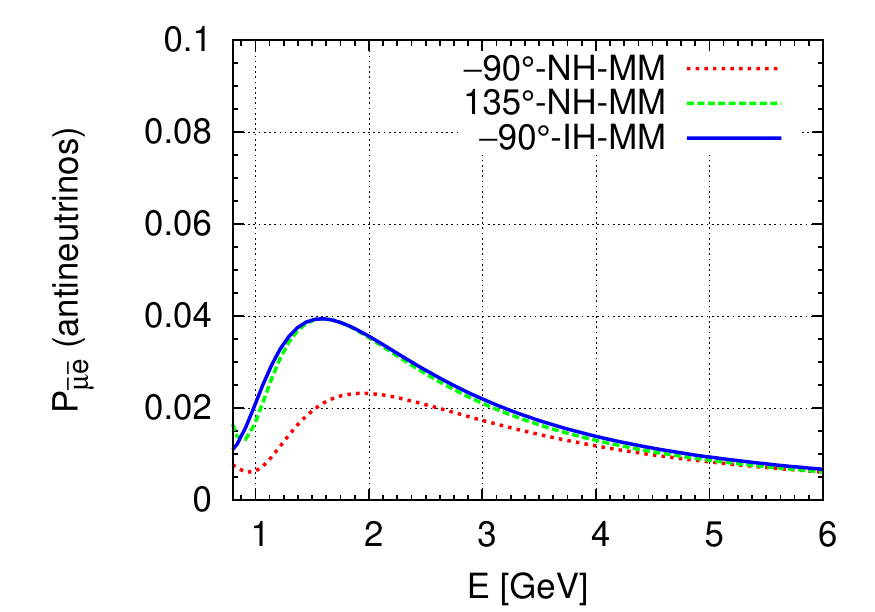}
		\\
		(a) Neutrinos.
		&
		(b) Antineutrinos.
	\end{tabular}
	\caption{Same as FIG. \ref{NOvAprob3nu}. but for $\theta_{23}=45^\circ$.\label{NOvAprob3nuMaxMix}}
\end{figure}

It can be seen that for the unprimed 3$\nu$ case (FIG. \ref{NOvAprob3nu}.), all three probability curves for \nova{} running neutrinos are almost entirely degenerate, though in the antineutrino case only the B and C solutions are degenerate. In the primed case (FIG. \ref{NOvAprob3nuMaxMix}.) the B$'$ and C$'$ solutions are distinct from the A$'$ solution for neutrinos and antineutrinos. Extending to 4$\nu$ shows bands that are also almost totally overlapping for neutrinos while for antineutrinos, 4$\nu$ the bands get closer together again but solution A is still mostly separate. For the primed solutions the A$'$ band is still mostly distinct but now has significant overlap in both neutrinos and antineutrinos. 

\begin{figure}[h]
\begin{tabular}{cc}
	\includegraphics[width=0.25\textwidth]{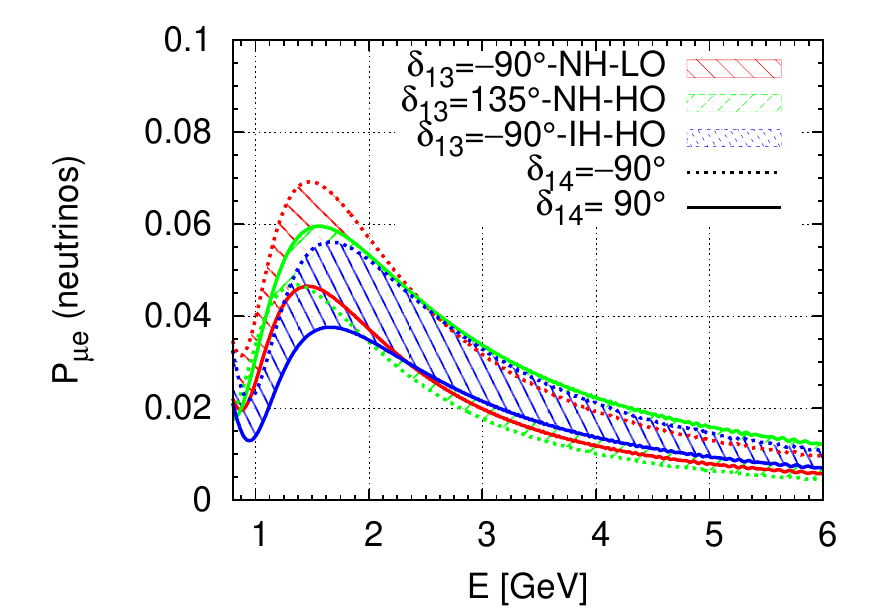}
	&
	\includegraphics[width=0.25\textwidth]{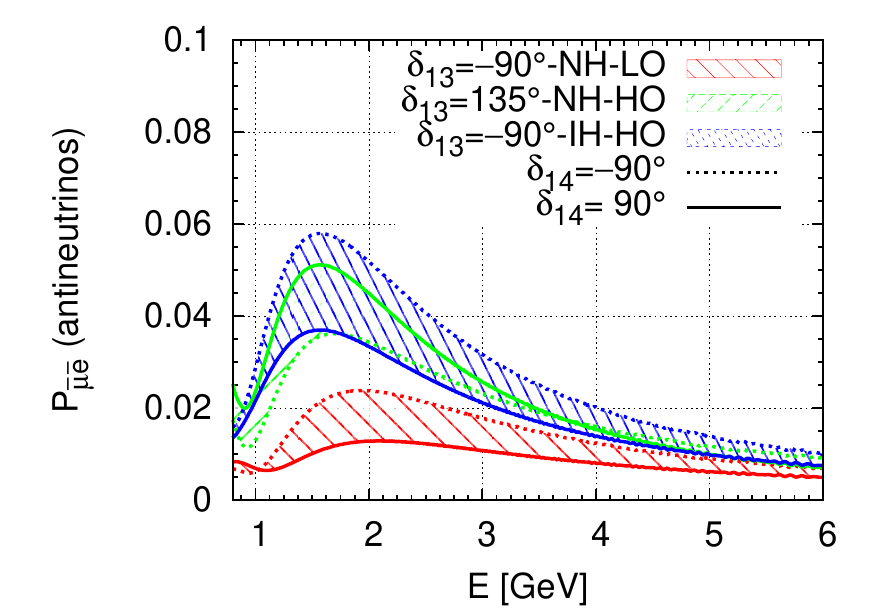}
	\\
	(a) Neutrinos.
	&
	(b) Antineutrinos.
\end{tabular}
\caption{Four-flavor probability plots with all three true value bands overlaid for \nova{}. The comparison between the neutrino and antineutrino cases is similar to the $3\nu$ case, but the LO and HO curves in the antineutrino case do get closer.\label{NOvAprob}}
\end{figure}

\begin{figure}[h]
	\begin{tabular}{cc}
		\includegraphics[width=0.25\textwidth]{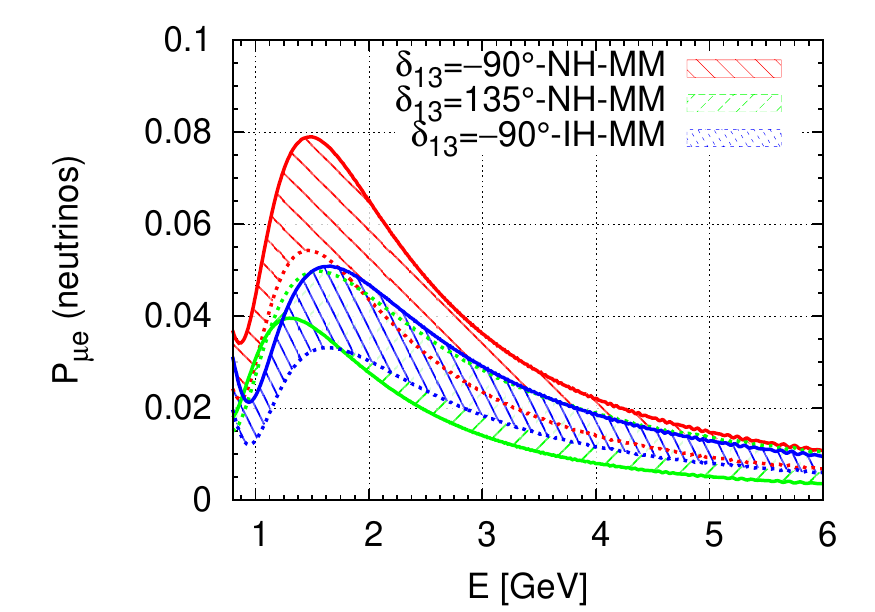}
		&
		\includegraphics[width=0.25\textwidth]{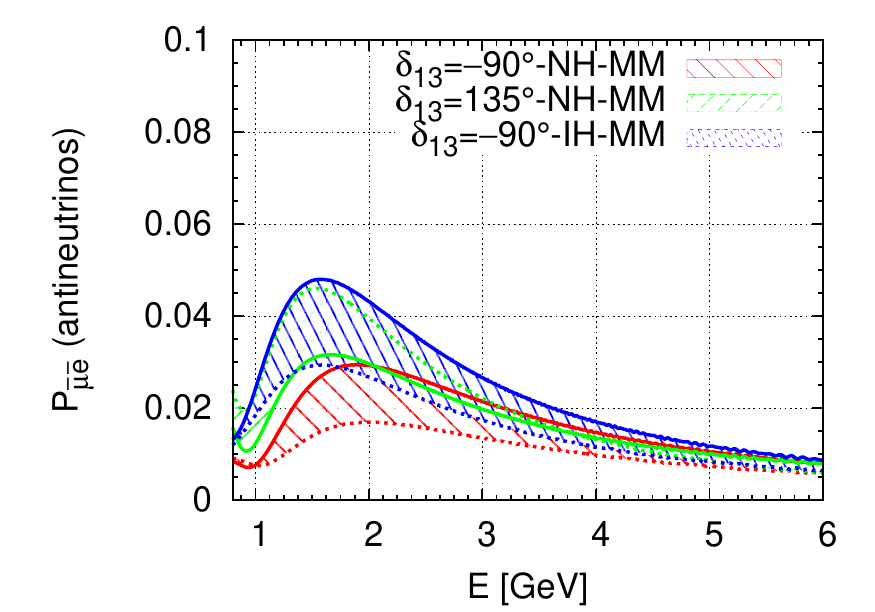}
		\\
		(a) Neutrinos.
		&
		(b) Antineutrinos.
	\end{tabular}
	\caption{Same as FIG. \ref{NOvAprob}. but for $\theta_{23}=45^\circ$.\label{NOvAprobMaxMix}}
\end{figure}

\subsubsection{DUNE}
\begin{figure}[h]
	\begin{tabular}{cc}
		\includegraphics[width=0.25\textwidth]{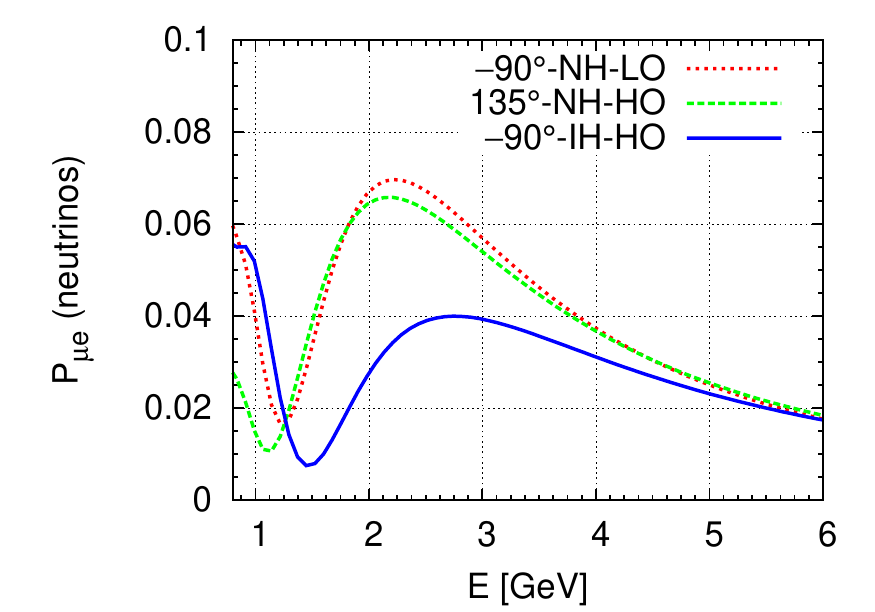}
		&
		\includegraphics[width=0.25\textwidth]{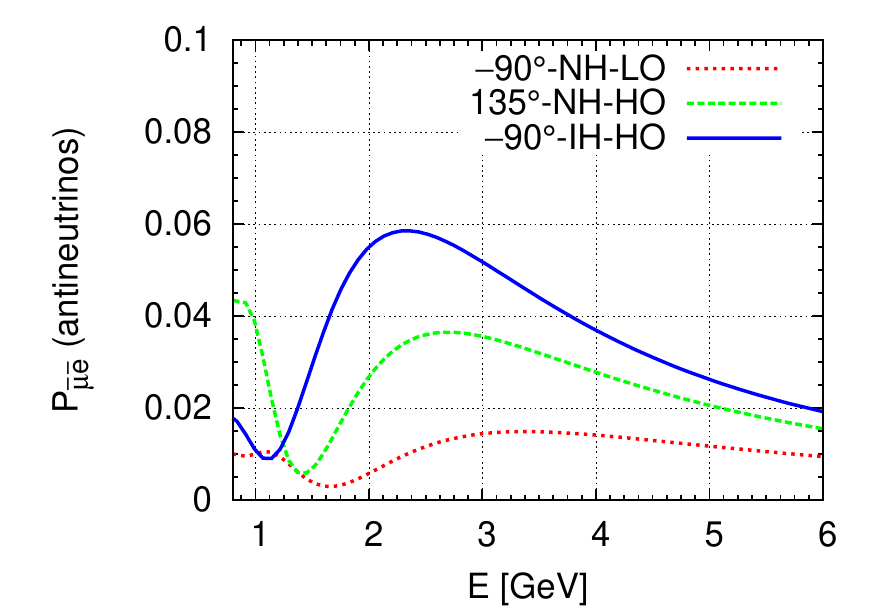}
		\\
		(a) Neutrinos.
		&
		(b) Antineutrinos.
	\end{tabular}
	\caption{Three-flavor probability plots with all three true value lines overlaid for DUNE, highlighting the larger separation of curves for the longer baseline detector.\label{DUNEprob3nu}}
\end{figure}

\begin{figure}[h]
	\begin{tabular}{cc}
		\includegraphics[width=0.25\textwidth]{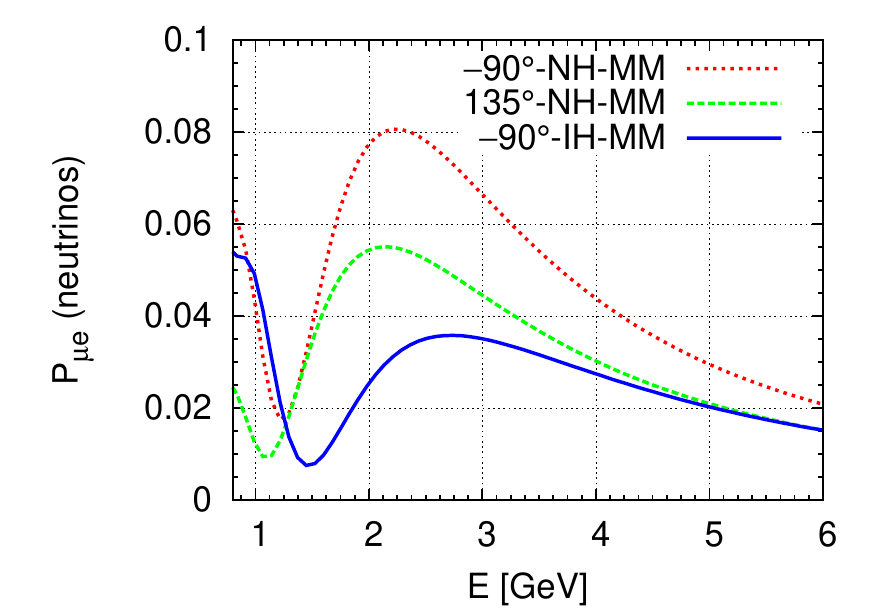}
		&
		\includegraphics[width=0.25\textwidth]{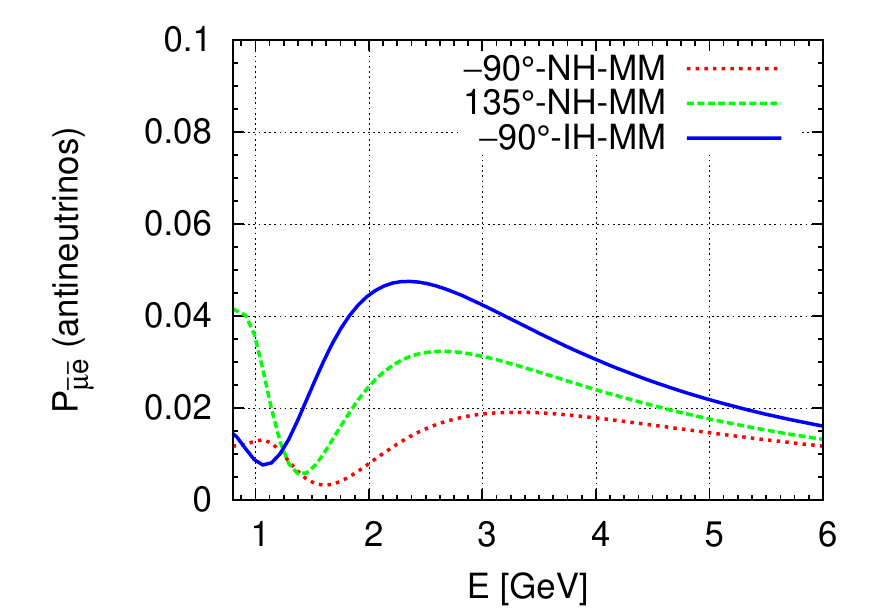}
		\\
		(a) Neutrinos.
		&
		(b) Antineutrinos.
	\end{tabular}
	\caption{Same as FIG \ref{DUNEprob3nu} but for $\theta_{23}=45^\circ$.\label{DUNEprob3nuMaxMix}}
\end{figure}

\begin{figure}[h]
	\begin{tabular}{cc}
		\includegraphics[width=0.25\textwidth]{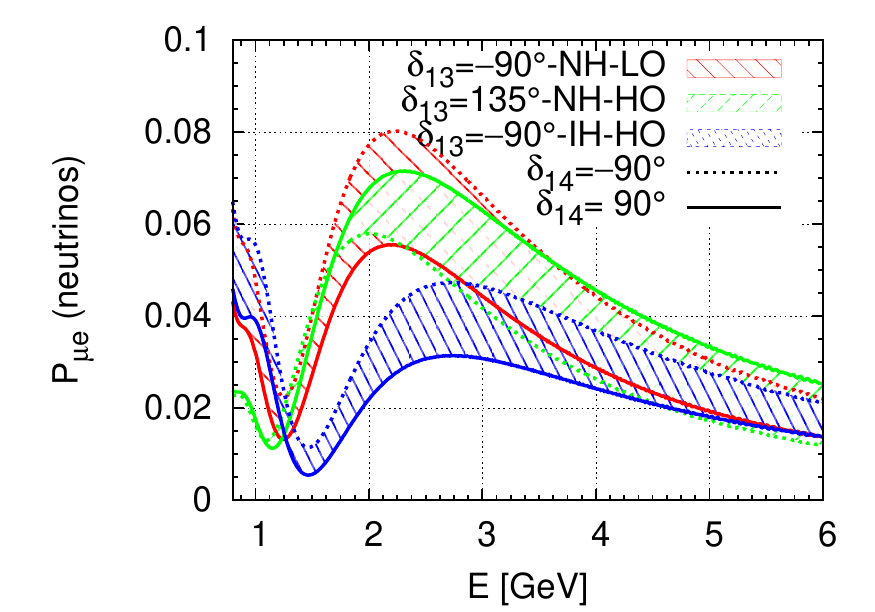}
		&
		\includegraphics[width=0.25\textwidth]{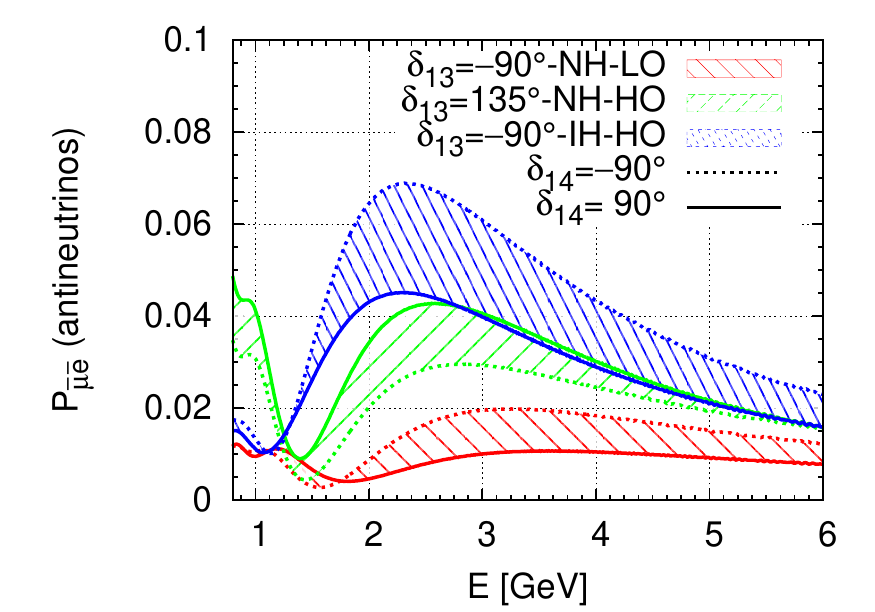}
		\\
		(a) Neutrinos.
		&
		(b) Antineutrinos.
	\end{tabular}
	\caption{Same as FIG. \ref{NOvAprob}. but for DUNE showing the minimal overlap introduced by the sterile CP phase $\delta_{14}$. \label{DUNEprob}}
\end{figure}

\begin{figure}[h]
	\begin{tabular}{cc}
		\includegraphics[width=0.25\textwidth]{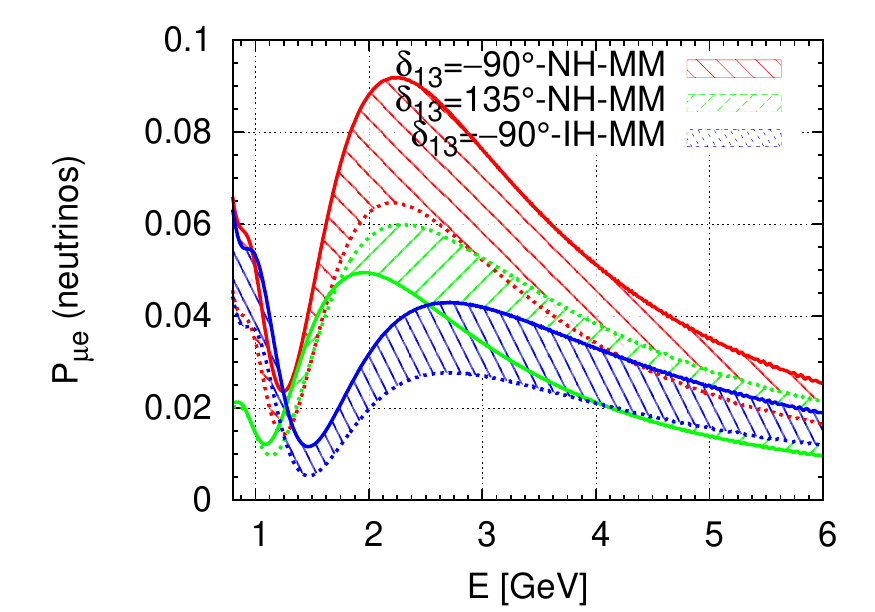}
		&
		\includegraphics[width=0.25\textwidth]{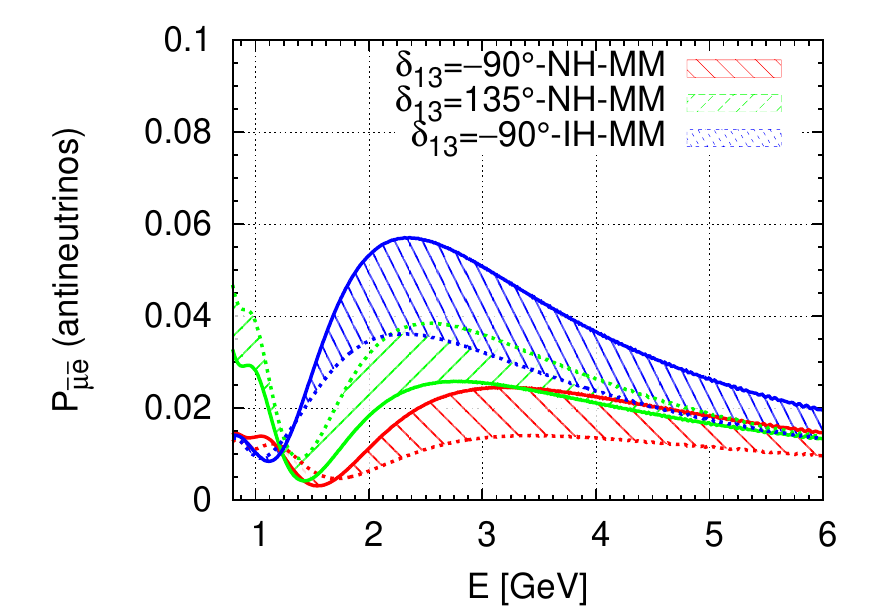}
		\\
		(a) Neutrinos.
		&
		(b) Antineutrinos.
	\end{tabular}
	\caption{Same as FIG \ref{DUNEprob} but for $\theta_{23}=45^\circ$. \label{DUNEprobMaxMix}}
\end{figure}

In contrast with the \nova{} plot, the 3$\nu$ DUNE plots (FIG. \ref{DUNEprob3nu}.) show only the A and B neutrino curves overlapping and no overlap for the antineutrino case, as shown in \cite{Goswami:2017hcw}. This points to much better degeneracy resolution than \nova{}, especially while running antineutrinos. The $4\nu$ plots (FIG. \ref{DUNEprob}.) do show overlap, specifically A, B and some C for neutrinos; and B and C for antineutrinos. Thus it is possible that some degeneracies can be reintroduced by extending our parameter space, even with the DUNE detector. Comparing these plots with the \nova{} ones shows that solution A is still the favored solution for degeneracy resolution. The probability plots do not tell the whole story however as they do not reflect the statistics of the detector, therefore we must do more analysis to get an idea of what significance degeneracies arise at. The primed MM case curves (FIG. \ref{DUNEprob3nuMaxMix}.) are widely spaced and have no overlap for DUNE in the 2-3GeV range. So if MM is the true case, DUNE should have better resolution power when running neutrinos and slightly worse power when running antineutrinos. Thus the MM case does not have a disparity in neutrino/antineutrino degeneracy resolution power unlike the octant cases. Similarly in the 4$\nu$ case (FIG. \ref{DUNEprobMaxMix}.) the neutrino overlap improves slightly, while the antineutrino overlap gets slightly worse.




\subsection{Degeneracies at the detector level}
We now analyse our test hypotheses using several $\chi^2$ type analyses to see for which values we can resolve the MH degeneracy, see what regions are allowed at 90\% C.L. and also to look at the CP sensitivity for a variety of true values. This is necessary because we need to account for statistical effects and  combined neutrino/antineutrino runs. Note that because $\theta_{34}$ and  $\delta_{34}$ do not come into the vacuum expression for $P_{\mu e}$ we set them to zero and do not marginalize. However, for neutrinos propagating in matter these extra mixing parameters will contribute from terms introduced by matter effects. Despite this, these contributions are small at \nova{} and DUNE and as such can be ignored when performing phenomenological analyses.

When performing the $\chi^2$ analysis we take the true parameters to be A, B or C (then A$'$, B$'$ and C$'$) and the test parameters to be as specified in TABLE \ref{StdParam} including marginalization ranges for the free parameters.

Our test statistic comes from GLoBES and is defined as:
\begin{equation}
\chi_{}^2
=
\sum_{i}^{}
\frac{\left(N_i^\mathrm{true}-N_i^\mathrm{test}\right)^2}
{N_i^\mathrm{true}}\,,
\end{equation}where $N_i^\mathrm{true}$ is the distribution for whatever the current true value is and $N_i^\mathrm{test}$ is the distribution for the test values that are varied over. This is calculated automatically by functions in by the GLoBES program with marginalization performed manually.

\subsubsection{NO$\nu$A}
\begin{figure}[h]
	\includegraphics[width=0.5\textwidth]{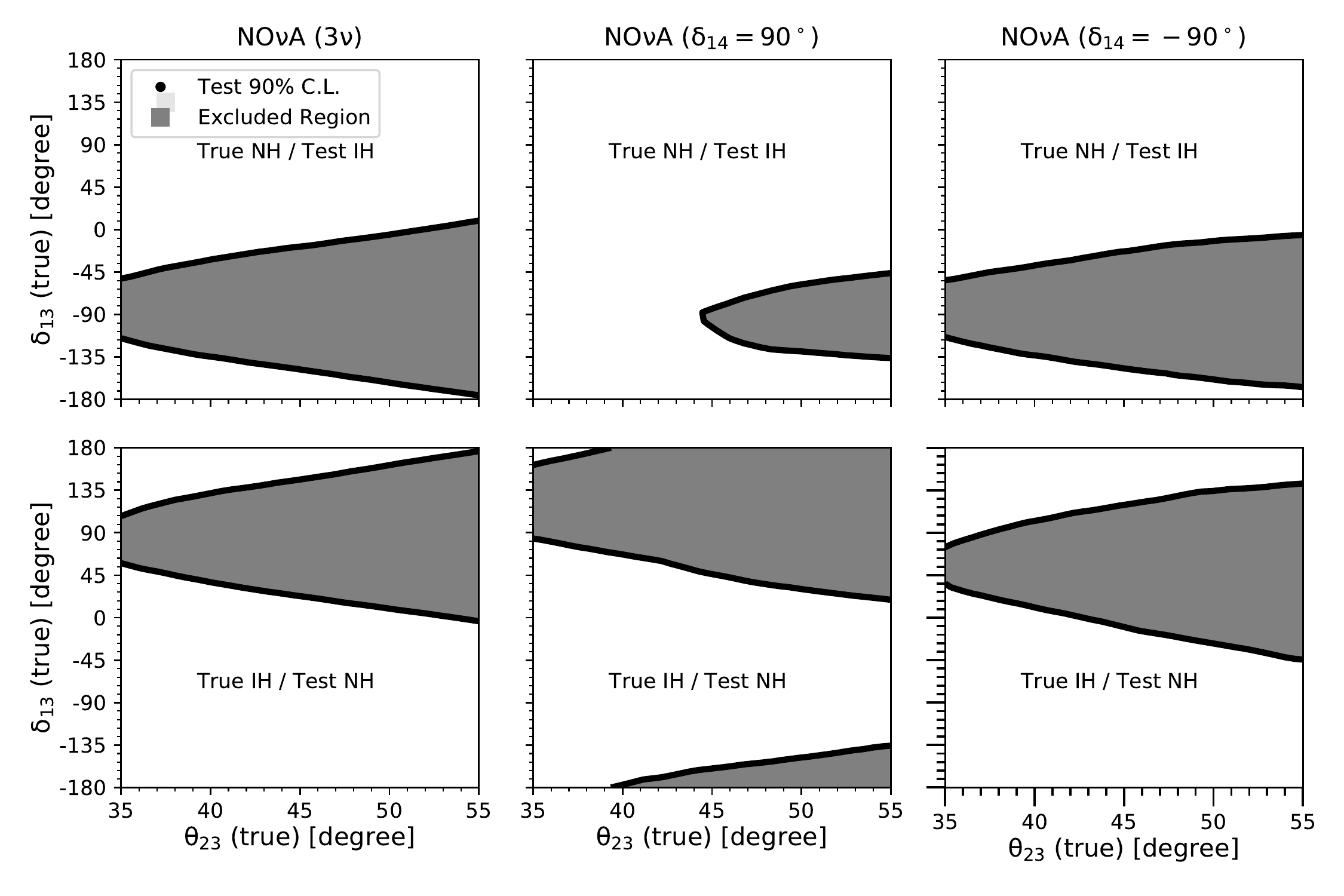}
	\caption{MH exclusion plots for \nova{} $\left(3+\bar{3}\right)$\label{exclNOvA}}
\end{figure}

\paragraph{Exclusion Plots}
To investigate the explicit range of true values for which the MH can be resolved we can create a new plot, known as a hierarchy exclusion plot, by varying the true oscillation parameters, flipping the hierarchy in the test hypothesis and marginalizing over every other variable. When we examine the exclusion plots for \nova{} (FIG. \ref{exclNOvA}.) we can see that the excluded region for true NH (true IH) includes the $\delta_{13}=+90^\circ$ ($\delta_{13}=-90^\circ$) favored region, this should be expected because for the favored parameters it is predicted that in the $3\nu$ case \nova{} alone can resolve the mass hierarchy. Extending into $4\nu$ changes these regions somewhat, e.g. for true NH, $\delta_{14}=90^\circ$ the exclusion zone retreats towards the HO side of our plot, indicating that the MH degeneracy can only be solved for true values roughly in the ranges: $\theta_{23}>45^\circ$ and $\delta_{13}\in(-45^\circ,-135^\circ)$. The change in the corresponding true IH plot with $\delta_{14}=90^\circ$ is much less extreme, still allowing MH resolution for some LO true values. On the other hand for $\delta_{14}=-90^\circ$ both NH and IH are mostly similar to the $3\nu$ case and as such the favored half planes are mostly excluded.

\begin{figure}[h]
	\includegraphics[width=0.5\textwidth]{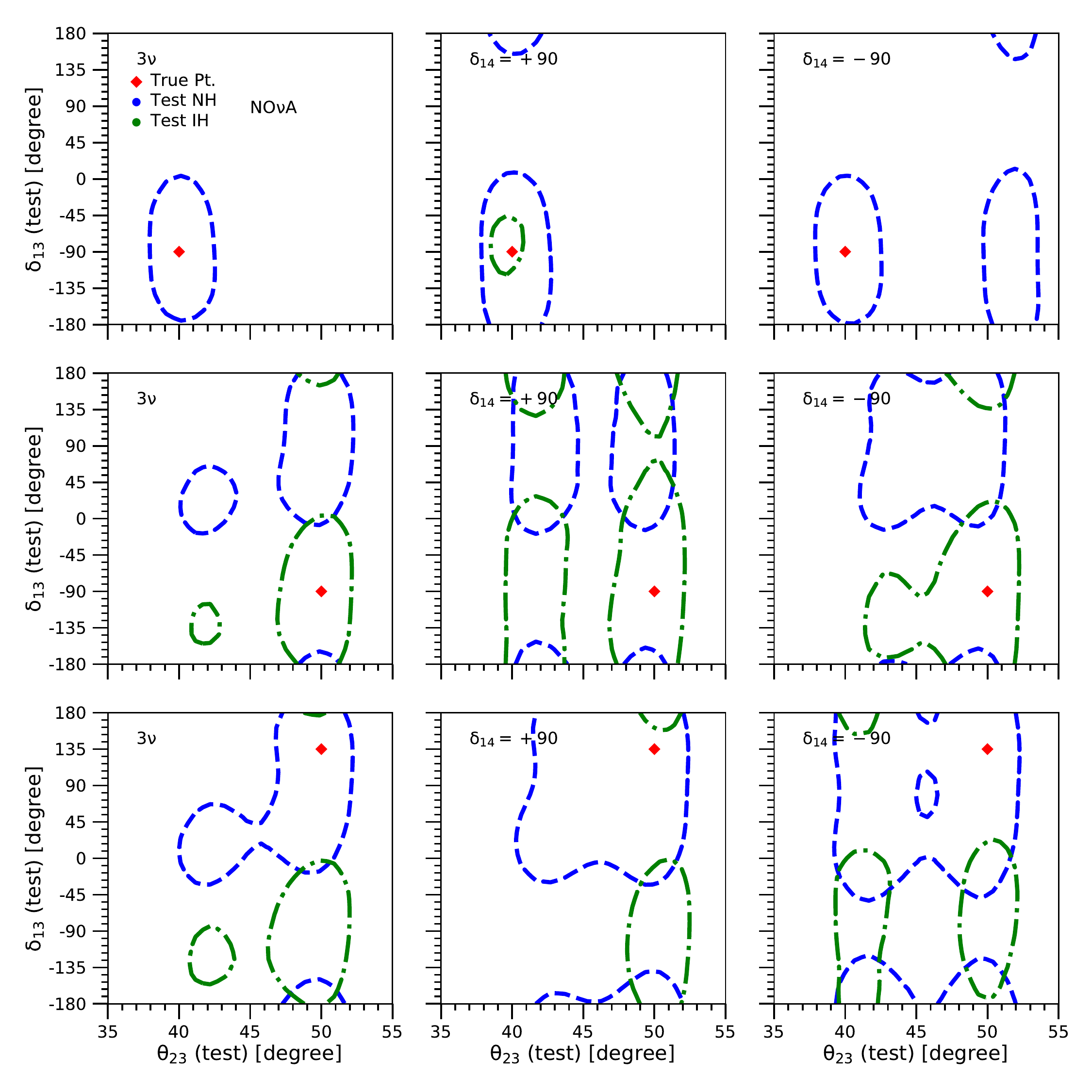}
	\caption{Allowed region plots in the test $\theta_{23}$-$\delta_{13}$ plane for three different true values of  $\delta_{13}$, $\theta_{23}$, $\mathrm{MH}$ for $3\nu$ (first column) as well as $\delta_{14}\pm90^\circ$ in $4\nu$ (second and third columns) all for \nova{}.\label{NOvAregions}}
\end{figure}

\paragraph{Allowed Region Plots}
From FIG. \ref{NOvAregions}. it can be seen that in the $3\nu$ case, the plot for A shows one allowed region surrounding the true value, while the B and C plots have WO-WH-W$\delta_{13}$, RO-WH-W$\delta_{13}$ and WO-RH-R$\delta_{13}$ regions as well as the correct solution. For the $4\nu$ cases, in general the regions are broadly the same, though for $\delta_{14}=+90^\circ$ true value A gains a WH region while for $\delta_{14}=-90^\circ$ it gains a WO region. More significantly, for true values B and C the regions mostly get larger (though the WO-WH-W$\delta_{13}$ solution for C vanishes). Overall FIG. \ref{NOvAregions}. shows that solution A can be resolved more easily than the other cases, by relating the probability plots to the allowed regions, the particularly large separation of the curves for antineutrinos compared to neutrinos contributes strongly to this.

\begin{figure}[h]
	\includegraphics[width=0.5\textwidth]{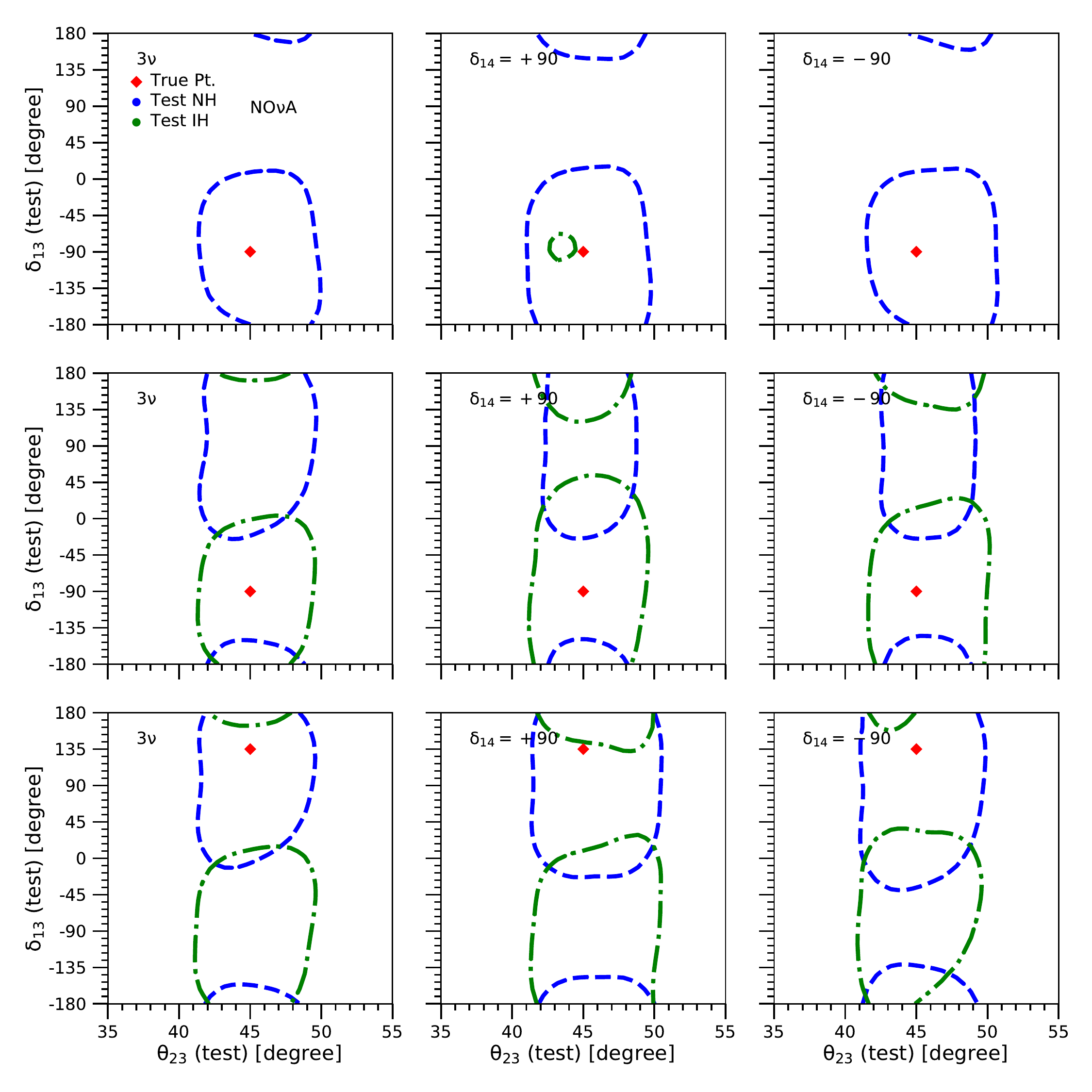}
	\caption{Same as FIG. \ref{NOvAregions}. but for $\theta_{23}=45^\circ$.\label{NOvAregionsMaxMix}}
\end{figure}

Similarly in FIG. \ref{NOvAregionsMaxMix}. the A$'$ case is still the one with the least degeneracy
having only a small WH solution when $\delta_{14}=+90^\circ$. In the other MM cases the MH degeneracy exists with regions almost reflected about $\delta_{13}=0^\circ$. For most of these cases the LO and HO solutions we tested ($\theta_{23}=40^\circ,50^\circ$) are just outside the 90\% C.L. regions, though $\theta_{23}\approx42.5^\circ,49^\circ$ are included in all regions, implying that some HO/LO solutions with less extreme values can't be ruled out by \nova{} in the MM case.

\subsubsection{DUNE}
\begin{figure}[h]
	\includegraphics[width=0.5\textwidth]{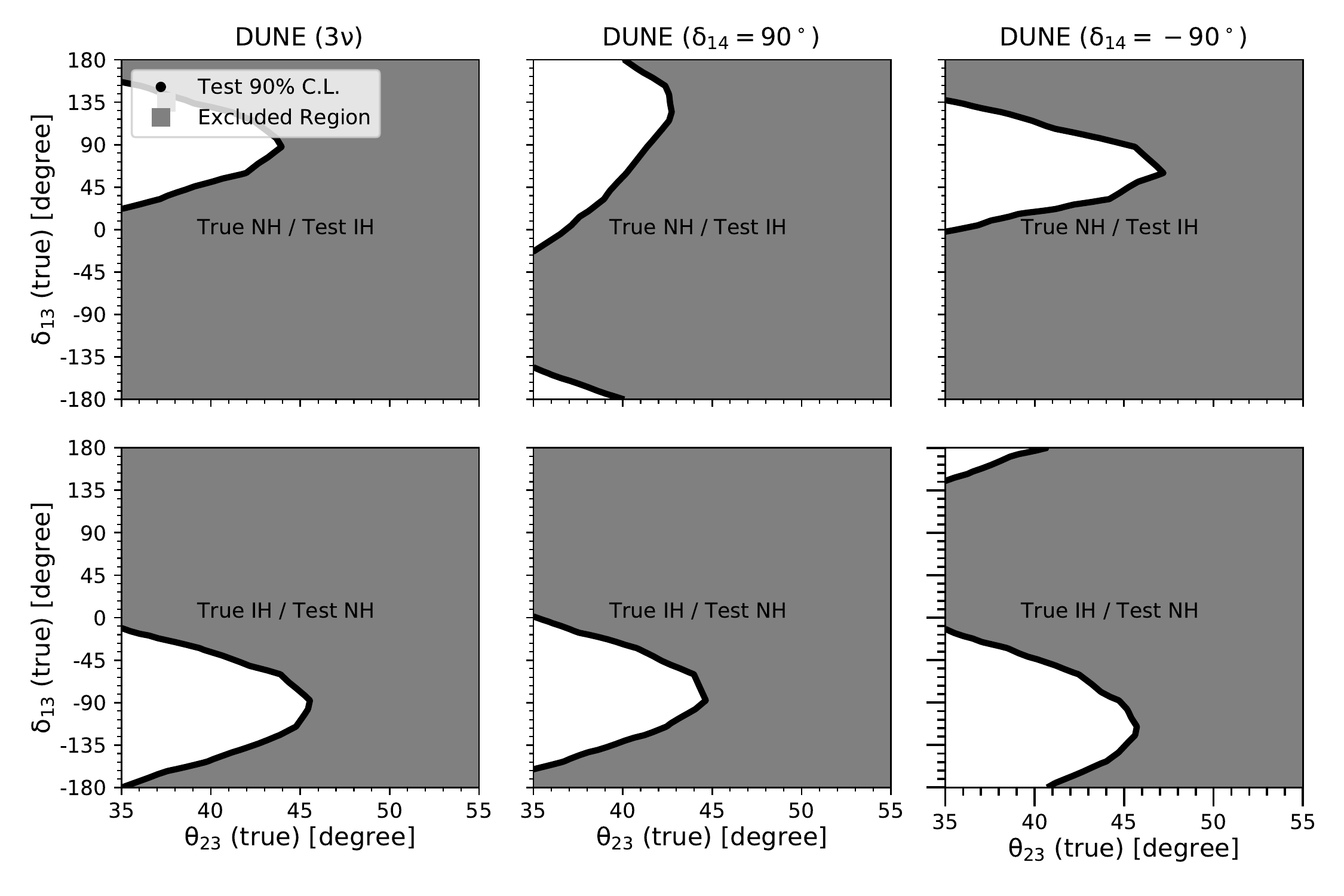}
	\caption{MH exclusion plots for DUNE $\left(2+\bar{2}\right)$\label{exclDUNE2+2}}
\end{figure}

\paragraph{Exclusion Plots}
Evaluating the exclusion plots for the reduced or partial run of DUNE $2+\bar{2}$ (FIG. \ref{exclDUNE2+2}.) and comparing to \nova{} shows that the excluded region expands to include much of the unfavored half plane. On the $\theta_{23}<45^\circ$ side of the plot there is a reasonable area still allowed implying that true LO is unfavored for degeneracy resolution, even at DUNE. In the $\delta_{14}=-90^\circ$ cases there is still a small spread at $\theta_{23}=45^\circ,\delta_{13}\approx\pm90^\circ$ for true MH=NH/IH, in which MH degenerate solutions will still exist.

\begin{figure}[h]
	\includegraphics[width=0.5\textwidth]{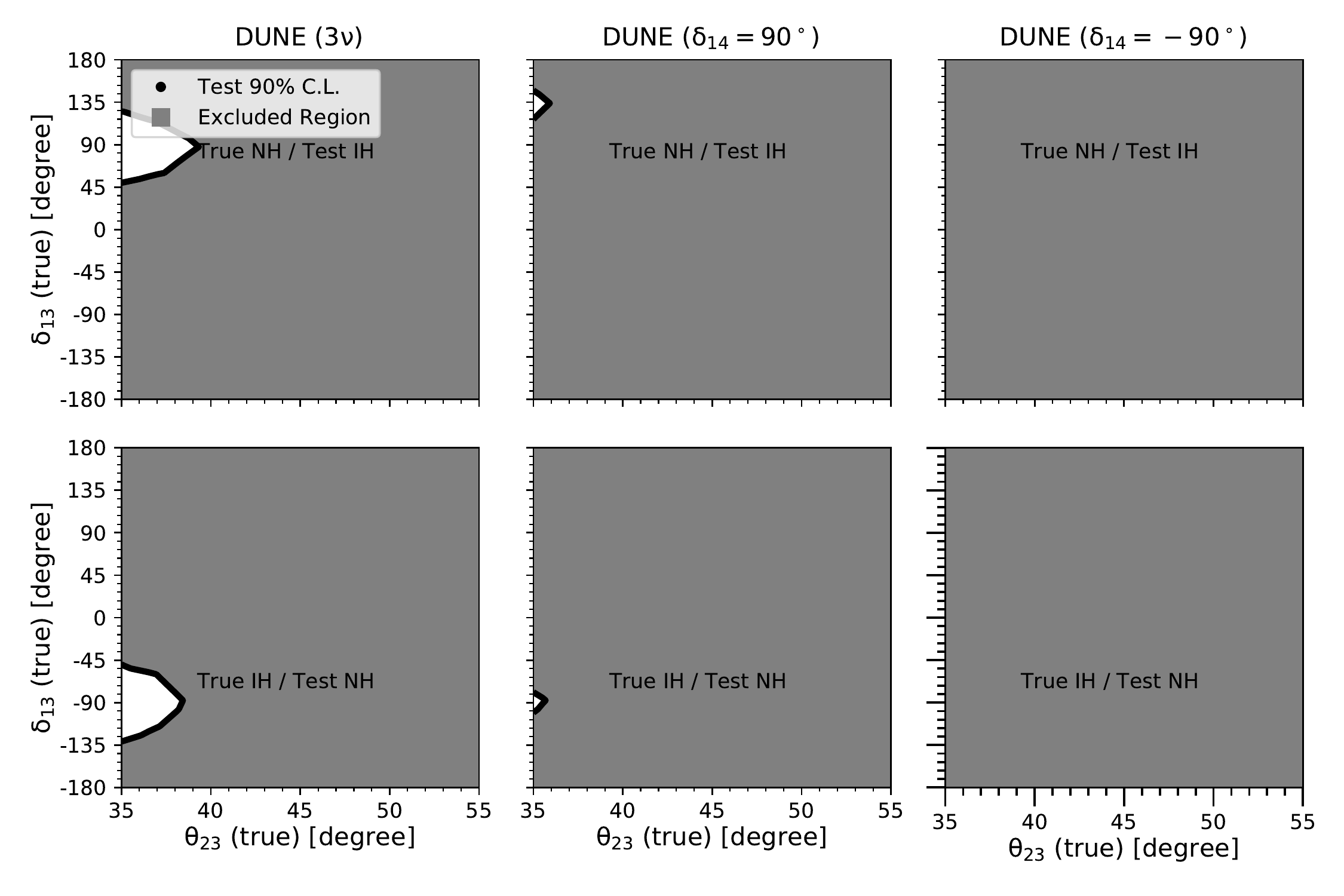}
	\caption{MH exclusion plots for DUNE $\left(5+\bar{5}\right)$\label{exclDUNE5+5}}
\end{figure}

Extending DUNE's run to $5+\bar{5}$ further increases the parameter space for which the wrong mass hierarchy can be excluded (FIG. \ref{exclDUNE5+5}.) and only small areas in the unfavored half-planes remain for $\theta_{23}<40^\circ$ which is roughly $2\sigma$ to $3\sigma$ outside of \nova{}'s current fits depending on the value of $|\Delta m^2_{31}|$. Because these non-excluded values are only valid for $\theta_{23}$ well below current LO estimates this reinforces the prediction that after it's full run, DUNE will be capable of resolving the MH degeneracy independently of other experiments, regardless of $\theta_{23}$, even in the case of small sterile mixing.

\begin{figure}[h]
	\includegraphics[width=0.5\textwidth]{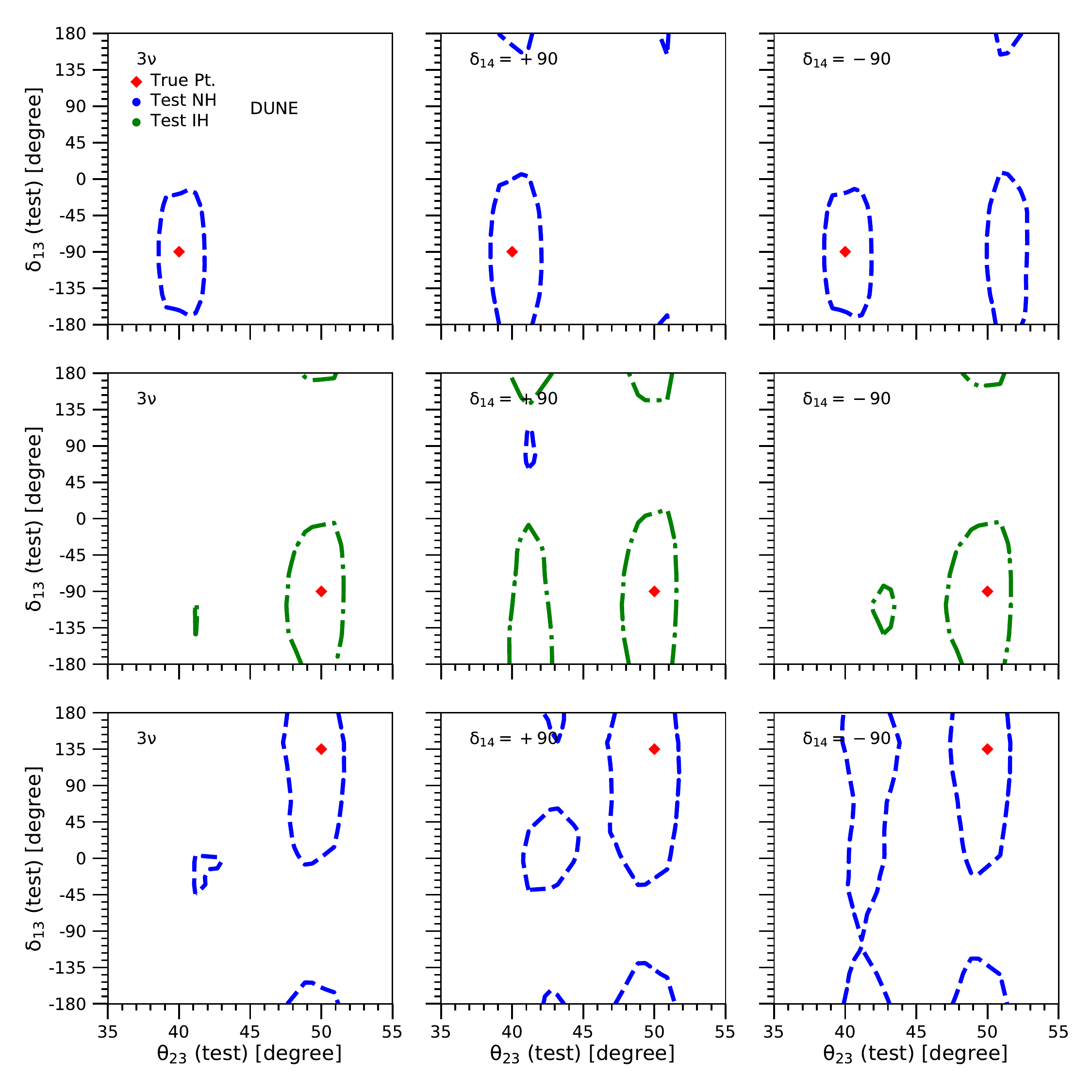}
	\caption{The same as FIG. \ref{NOvAregions}. but for DUNE $\left(2+\bar{2}\right)$\label{DUNEregions2+2}.}
\end{figure}

\paragraph{Allowed Region Plots}
Evaluating the allowed regions for DUNE $2+\bar{2}$ shows an almost complete disappearance of WH solutions. Many of the WO solutions are gone too, for example the $3\nu$ IH scenario in FIG. \ref{DUNEregions2+2}. Though some cases are still particularly bad, e.g. also in FIG. \ref{DUNEregions2+2}.; true value B, which has a degenerate octant solution that almost spans $\delta_{13}$'s entire range.

\begin{figure}[h]
	\includegraphics[width=0.5\textwidth]{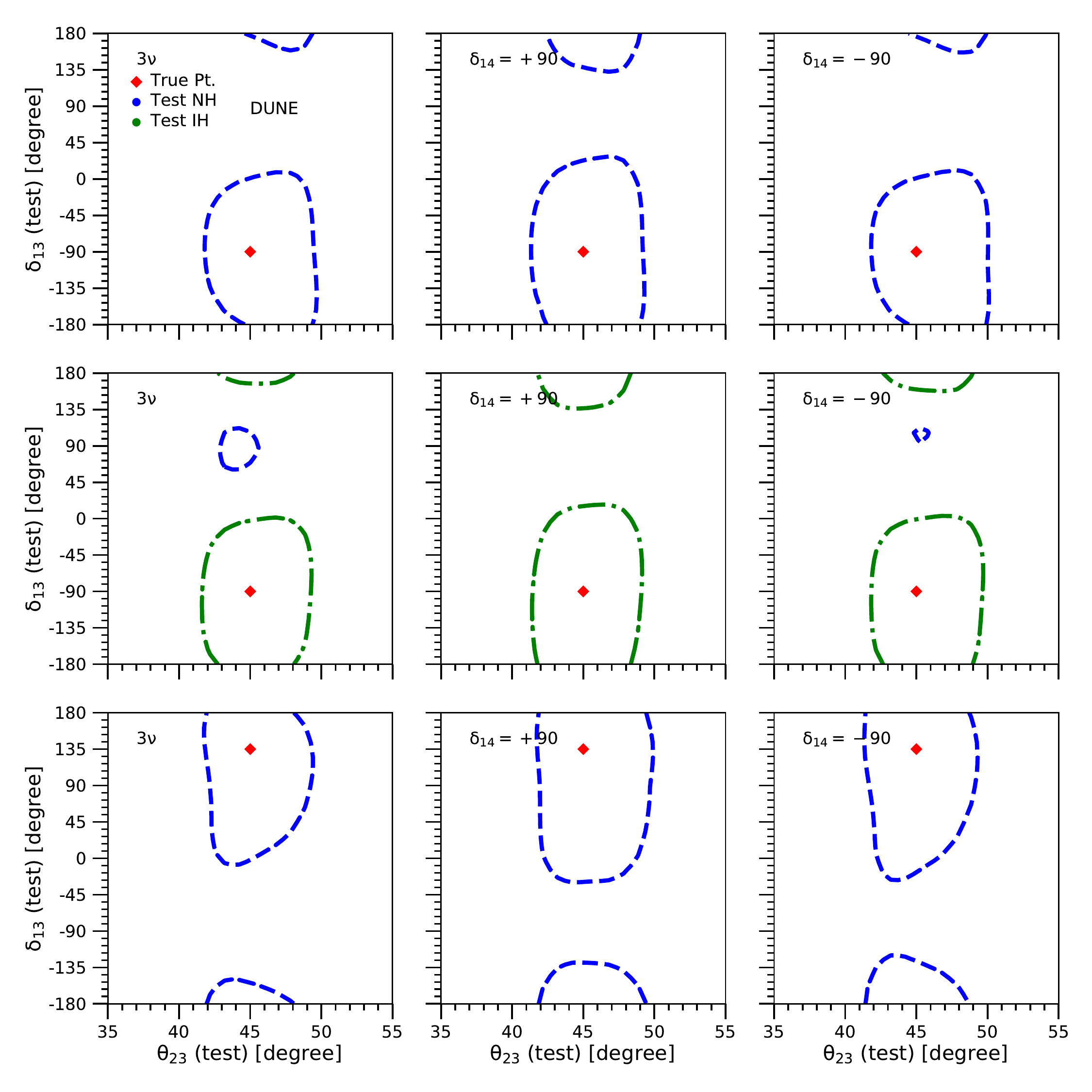}
	\caption{Same as FIG. \ref{DUNEregions2+2}. but for $\theta_{23}=45^\circ$\label{DUNEregions2+2MaxMix}.}
\end{figure}

For the MM case with only $2+\bar{2}$ running (FIG. \ref{DUNEregions2+2MaxMix}.) the MH degenerate regions present for \nova{} vanish for most cases and only remain for B$'$ $3\nu$ and $\delta_{14}=-90^\circ$ as small regions. The size of the regions does not change much compared to  \nova{} so the allowed $\theta_{23}$ range is roughly the same though the allowed regions do avoid $\theta_{23}=40^\circ,50^\circ$ in more of the cases. Overall for DUNE ($2+\bar{2}$) the trade off is between octant true values with degenerate solutions or max-mixing true values with more uncertainty in the exact value of $\theta_{23}$.

\begin{figure}[h]
	\includegraphics[width=0.5\textwidth]{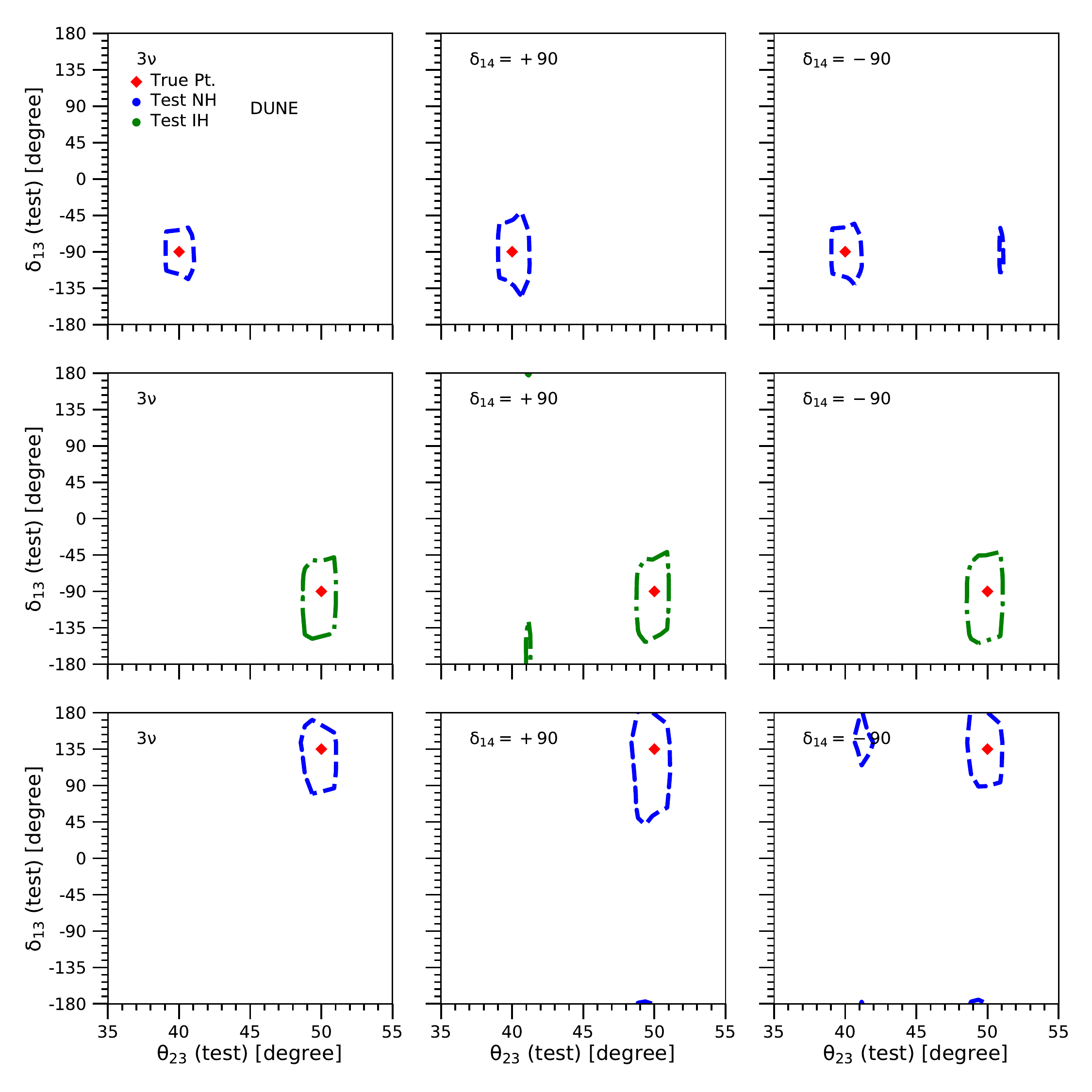}
	\caption{The same as FIG. \ref{NOvAregions}. but for DUNE $\left(5+\bar{5}\right)$\label{DUNEregions5+5}.}
\end{figure}

From FIG. \ref{DUNEregions5+5}. it can be seen that despite the additional probability overlap induced by the sterile parameters, for DUNE $5+\bar{5}$ the degeneracies are practically resolved at 90\% C.L. aside from small wrong octant regions for values A and C with $\delta_{14}=-90^\circ$ and for B with $\delta_{14}=+90^\circ$. This is due to the fact that hierarchy resolution ability is related to the baseline of the experiment and as seen in FIG. \ref{DUNEprob}. (b) at 2.5 GeV neutrino energy, DUNE has no overlap for our three parameter bands when running antineutrinos, this allows excellent degeneracy resolution. 

\begin{figure}[h]
	\includegraphics[width=0.5\textwidth]{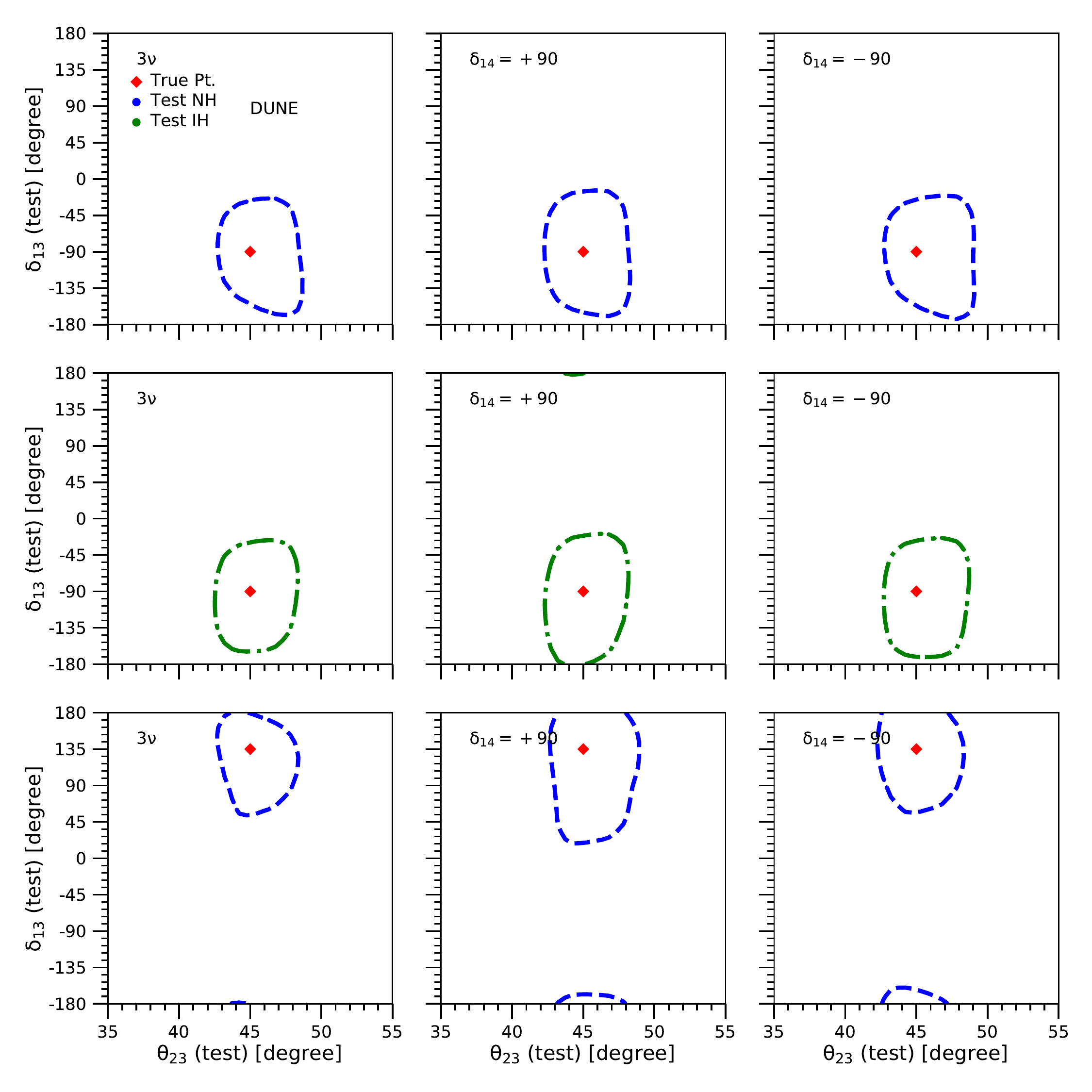}
	\caption{Same as FIG. \ref{DUNEregions5+5}. but for $\theta_{23}=45^\circ$\label{DUNEregions5+5MaxMix}.}
\end{figure}

In the MM case (FIG. \ref{DUNEregions5+5MaxMix}.) the allowed regions for DUNE get larger but have no MH degenerate regions. In all cases the HO/LO solutions are outside the 90\% C.L. regions implying good rejection of HO/LO solutions and a good contribution to the precision measurement of $\theta_{23}$.

\section{Conclusion}
We extend the analysis from \cite{Ghosh2017} in light of the discussions from \cite{Goswami:2017hcw} regarding the results in \cite{Adamson:2017gxd}. We include a light sterile neutrino specified as such to rectify the short baseline oscillation anomalies. From our analysis we see that the degenerate solutions are predicted to be worse at probability level for the 4$\nu$ case due to the additional free parameter space. We find that for certain values of $\delta_{14}$ the sensitivity of \nova{} to the octant degeneracy and (to a much lesser extent) hierarchy degeneracy may be reduced. We also predict that DUNE $2+\bar{2}$ can solve the MH degeneracy at 90\% C.L. while some octant ambiguity still exists. However, extending to the full DUNE $5+\bar{5}$ run removes almost all ambiguity at 90\% C.L. in all cases regardless of $\delta_{14}$. So it can be seen that for any of these true values with the sterile hypothesis being correct or not, that DUNE can resolve these degeneracies at 90\% C.L. whilst \nova{} alone loses some potential for degeneracy resolution in the sterile case.

We also find that if the $\theta_{23}$ value chosen by nature is $45^\circ$, then the need for combined neutrino/antineutrino analysis to distinguish certain results is diminished. This leads to increased MH resolution power but less precision for the exact value of $\theta_{23}$. However it can be seen that DUNE has similar MH resolution power at 90\% C.L. no matter the case. It remains to be seen over the next few years how important DUNE will be in this field, depending on what best fit parameters \nova{} and T2K favor.

\section{Additional Notes}
\begin{figure}[b]
	\includegraphics[width=0.5\textwidth]{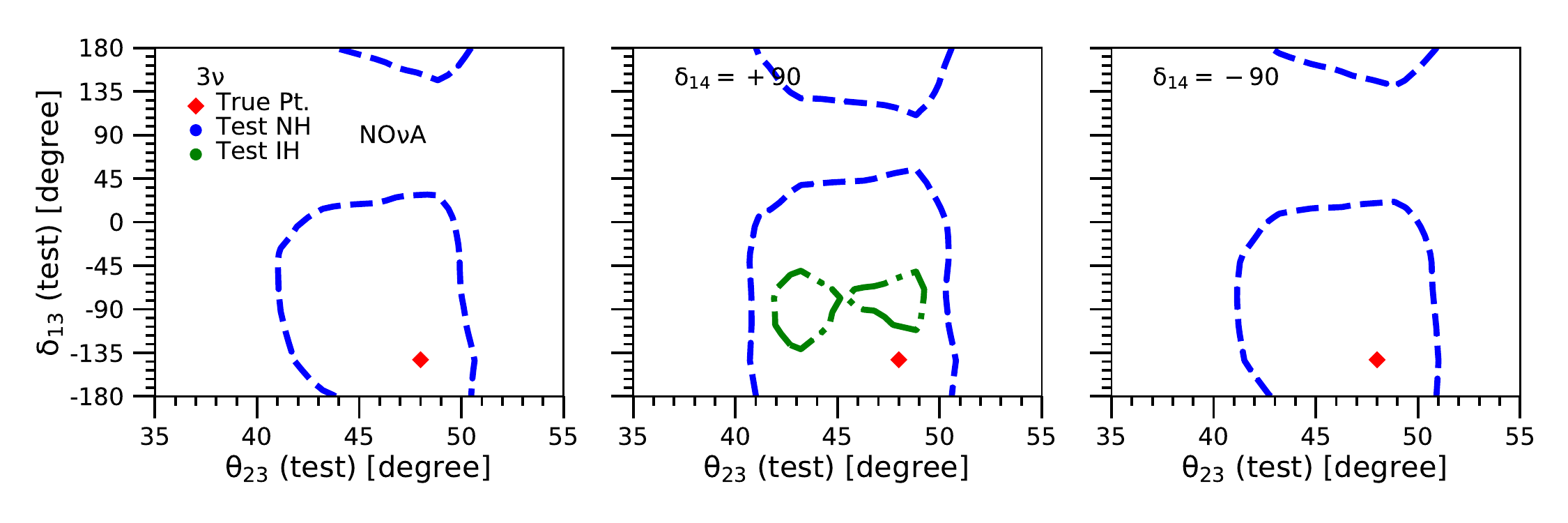}
	\caption{Allowed regions for the new preliminary best fits for \nova{} $\left(3+\bar{3}\right)$ with $4\nu$ extension\label{newNOvA}.}
	\includegraphics[width=0.5\textwidth]{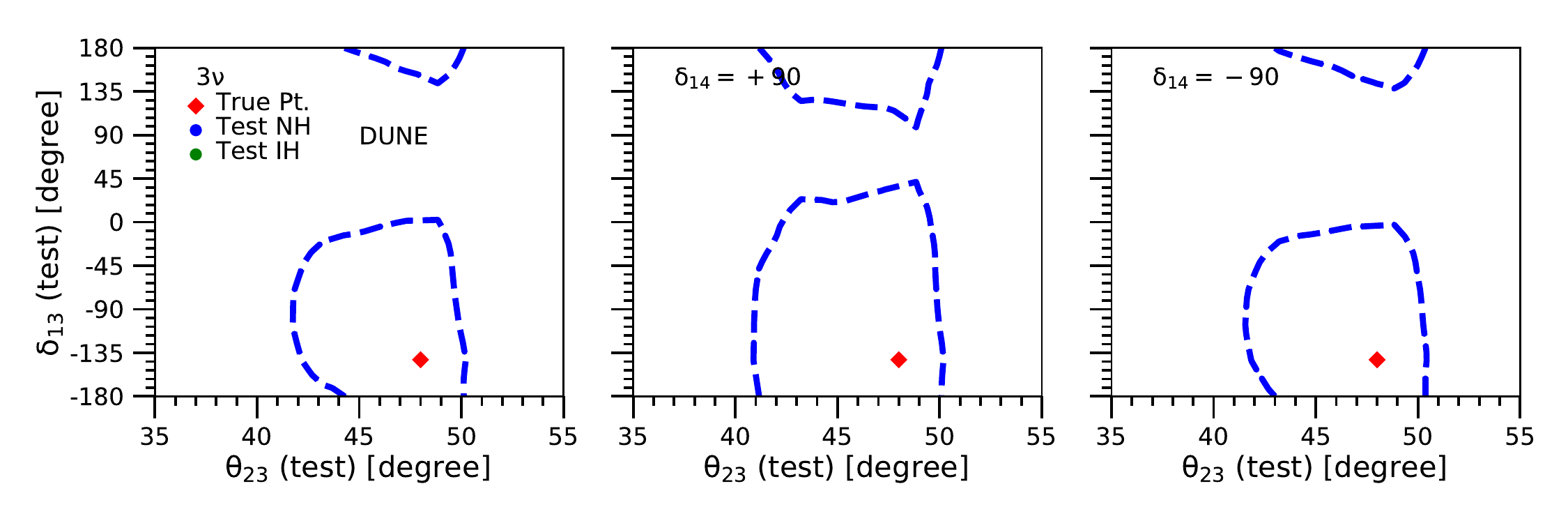}
	\caption{The same as FIG. \ref{newNOvA}. but for DUNE $\left(2+\bar{2}\right)$\label{newDUNE2+2}.}
\end{figure}

New results from NOvA have been published recently \cite{NOvA:2018gge,Radovic:2018,Himmel:2018} 
and indicate new $1\sigma$ parameter ranges:
\begin{align}
\Delta m^2_{32}
&=
2.444^{+0.079}_{-0.077}\times 10^{-3}\mathrm{eV}^2\\
\sin^2\theta_{23}
&=
\begin{cases}
0.558^{+0.041}_{-0.033} \left(\mathrm{HO}\right)\\
0.475^{+0.036}_{-0.044} \left(\mathrm{LO}\right)
\end{cases}
\end{align}with best fits of: $\delta_{13}=1.21\pi\approx-142.2^\circ$, HO, NH. These align somewhat better with previous T2K and MINOS results and no-longer explicitly rule out $\theta_{23}=45^\circ$ at 90\% C.L. We will still continue to analyse our three values despite the fact that neither A or B are fully favored and C is disfavored, because we are interested purely in degeneracy resolution. With regards to these new preliminary best fits from \nova{}, our sensitivity predictions do not really change, these results still fall into the favored area for mass hierarchy resolution and as such the \nova{} only loses MH sensitivity in the specific $4\nu$ case with $\delta_{14}=-90^\circ$ (FIG. \ref{newNOvA}.). The octant region does have more spread for this true value, but the allowed region doesn't include the wrong octant, instead including the maximal-mixing ($\theta_{23}=45^\circ$) case. For DUNE $\left(2+\bar{2}\right)$ the results are similar (FIG. \ref{newDUNE2+2}.). Therefore in this case MM can not be ruled out at 90\% C.L. and may require a combined analysis to differentiate.
\label{sec4}

\section{Acknowledgement}
SG, ZMM, PS and AGW thank the support by the University of Adelaide and the Australian Research Council through the ARC Centre of Excellence for Particle Physics (CoEPP) at the Terascale (grant no. CE110001004).

\end{document}